\documentclass[%
 reprint,
 floatfix,
 nofootinbib,
 amsmath,amssymb,
 aps,
]{revtex4-1}

\usepackage{graphicx}
\usepackage{dcolumn}
\usepackage{bm}
\usepackage{upgreek}
\usepackage{hyperref}
\usepackage{cleveref}
\usepackage{xcolor}
\usepackage{ae,aecompl}
\usepackage{graphicx}
\usepackage{amsmath}
\usepackage{amssymb}
\usepackage{txfonts}
\usepackage{bm}
\usepackage{tensor}
\usepackage{mathtools}

\newcommand{\ltsima}{$\; \buildrel < \over \sim \;$}
\newcommand{\lsim}{\lower.5ex\hbox{\ltsima}}
\newcommand{\gtsima}{$\; \buildrel > \over \sim \;$}
\newcommand{\gsim}{\lower.5ex\hbox{\gtsima}}

\begin{document}

\preprint{APS/123-QED}

\title[PNGs with FRBs]
{Probing primordial non-Gaussianity with Fast Radio Bursts}

\author{Robert Reischke}
  \email{r.reischke@campus.technion.ac.il}
\affiliation{%
Department of Natural Sciences, The Open University of Israel, 1 University Road, P.O. Box 808, Ra'anana 4353701, Israel\\ 
Department of Physics, Technion, Haifa 32000, Israel}

\author{Steffen Hagstotz}
\affiliation{
 The Oskar Klein Centre for Cosmoparticle Physics,  Department of Physics, Stockholm University, Roslagstullsbacken 21A, SE-106 91 Stockholm, Sweden
}%

\author{Robert Lilow}
\affiliation{%
 Department of Physics, Technion, Haifa 32000, Israel
}%

\date{\today}

\begin{abstract}
Fast radio bursts (FRBs) are astrophysical transients of currently unknown origin, and so far several events have been detected at extragalactic distances. The dispersion measure (DM) of the radio signal is a probe of the integrated electron density along the line of sight and therefore allows to map the electron distribution within the large-scale structure. Since a fraction of electrons gets expelled from galaxies by feedback, they are anticorrelated with halos at large scales and hence the angular DM correlations show a scale-dependent bias caused by primordial non-Gaussianity. Although the signal is weaker than in other probes like galaxy clustering, FRBs can potentially probe considerably larger volumes. We show that while studying the FRB clustering signal requires very large samples, correlations in the DM are cosmic-variance limited on large angular scales with only $\sim 10^{3-4}$ events. A tomographic analysis of the angular DM correlation function can constrain the local primordial bispectrum shape parameter $f_\mathrm{NL}$ to a precision down to ${f_\mathrm{NL}}\sim \mathcal{O}(1)$ depending on assumptions about the FRB redshift distribution and the astrophysical feedback on large scales. This makes FRBs a competitive probe to constrain inflationary physics.
\end{abstract}

\maketitle

\section{Introduction}
Fast radio bursts (FRBs) are very short transients, lasting typically a few milliseconds, with frequencies ranging from $\sim 100$ MHz to several GHz. Since the radio signal is travelling through the ionized intergalactic medium (IGM), each frequency of the pulse experiences a delay characterised by the dispersion measure (DM) proportional to the integrated free electron density along the line-of-sight \citep[e.g.][]{thornton_population_2013, petroff_real-time_2015, connor_non-cosmological_2016, champion_five_2016,chatterjee_direct_2017}. The mechanism responsible for the radio emission is currently unknown,\footnote{For a compilation of currently proposed mechanisms for FRBs, see \href{https://frbtheorycat.org/index.php/Main_Page}{https://frbtheorycat.org/index.php} \citep{platts_living_2019}.} but the isotropic occurrence of the events detected so far shows no alignment with the Milky Way disk and the measured DM for most events is very large, suggesting an extragalactic origin. In general, the total DM associated with an extragalactic FRB event consists of contributions from the host galaxy, the Milky Way and the diffuse electrons in the large-scale structure (LSS). Several authors therefore proposed to use the DM inferred from FRBs as a cosmological probe using either the averaged signal \citep{zhou_fast_2014,walters_future_2018} or the statistics of DM fluctuations \citep{masui_dispersion_2015,shirasaki_large-scale_2017,rafiei-ravandi_characterizing_2019}.

Currently more than $\sim100$ FRBs are detected\footnote{See \href{http://frbcat.org/}{http://frbcat.org/} for a catalog of known events \citep{petroff_frbcat_2016}}, but surveys such as CHIME, UTMOST, HIRAX, ASKAP or SKA \citep{johnston_science_2008, maartens_cosmology_2015, caleb_fast_2016,newburgh_hirax_2016} aim to provide $\sim 10^4$ FRBs per decade. Although FRBs do not show spectral features that allow for accurate redshift estimation, the accumulated DM can be translated into a noisy distance estimate \citep{masui_dispersion_2015}. Since FRBs are observed with $\mathrm{DM} \geq 1000\;\mathrm{pc}\;\mathrm{cm}^{-3}$ and models of the Milky way suggest $\mathrm{DM} \leq50\;\mathrm{pc}\;\mathrm{cm}^{-3}$ due to gas in our Galaxy for most directions on the sky \citep{yao_new_2017}, the majority of the DM signal originates from the diffuse electrons in the IGM, which traces the LSS on large scales. Although the host galaxy contribution is still under debate, its magnitude is expected to be comparable to that of the Milky Way.

Measurements of the DM are therefore a probe of the cosmic electron density.
Since electrons get expelled from structures by {astrophysical} feedback, they are anticorrelated with halos \citep{shaw_deconstructing_2012}. Effects that change the halo bias at large scales are therefore transferred to the clustering properties of the diffuse electron component via astrophysical feedback.

It has been long understood that the bias of halos with respect to the overall matter distribution \citep[see][for a review]{desjacques_large-scale_2018} opens a window to probing inflationary physics via a characteristic scale-dependent imprint induced by primordial non-Gaussianities (PNGs).
A small amount of PNGs is a prediction of various different inflationary models \citep[e.g.][]{bartolo_non-gaussianity_2004,komatsu_hunting_2010} and its detection could shed light on the physics of the very early Universe.

 Precise constraints on PNGs from any measurements are very sensitive to the effective survey volume, since the scale-dependent bias is proportional to $k^{-2}$ \citep{carbone_non-gaussian_2008, carbone_cosmological_2010}. Since the large scales in the cosmic microwave background (CMB) are already measured, future CMB surveys are unlikely to improve the constraints provided by \citet{planck_collaboration_planck_2018-1}, and any progress must come from the large-scale structure. Upcoming LSS observations promise to improve current constraints by roughly one order of magnitude by measuring the galaxy power spectrum at very large scales \citep{raccanelli_measuring_2015,camera_cosmology_2015,alonso_ultra-large-scale_2015,raccanelli_cosmological_2016}. FRBs are observable over almost the full sky and up to cosmological distances, so they have the potential to probe large volumes comparable to those probed by next-generation galaxy surveys. They are also less affected by foregrounds that dominate the signal, as for example is the case for the cosmic infrared background \citep{tucci_cosmic_2016}.

In this paper we investigate the potential to constrain PNGs using the tomographic DM angular power spectrum. We focus on the local type of PNG with the degree of non-Gaussianity described by the proportionality constant $f_\mathrm{NL}$. 
The paper is organized as follows: In \cref{sec:modelling_frb} we describe our model of the angular FRB statistics including primordial non-Gaussianity and the survey characteristics used for our forecast. In \cref{sec:results} we present the results and then summarize our findings in \cref{sec:conclusion}.

Since we are mainly interested in the effect of PNG on the large-scale bias, which cannot be produced by other cosmological parameters, {we} adopt a fiducial $\Lambda$CDM cosmology with $f_\mathrm{NL}=0$ and
all other parameters fixed to the respective best-fit values measured by \citet{planck_collaboration_planck_2018}.

\section{FRB statistics}
\label{sec:modelling_frb}

\subsection{Dispersion measure}
\label{subsec:dispersion_measure}
The frequencies of the FRB signal undergo dispersion as they travel through the ionized intergalactic medium. This causes a frequency-dependent offset between the arrival time of the different components of the pulse. 
The time delay measured from an FRB at redshift $z$ in direction $\hat{\boldsymbol{x}}$ can be written as
\begin{equation}
\label{eq:delay_measure}
    \delta t(\hat{\boldsymbol{x}},z) \propto \mathrm{DM}_\mathrm{tot}(\hat{\boldsymbol{x}}, z) \nu^{-2}\;,
\end{equation}
with the total dispersion measure $\mathrm{DM}_\mathrm{tot}(\hat{\boldsymbol{x}},z)$ proportional to the column density of electrons integrated along the line-of-sight. It can be broken up into individual contributions
\begin{equation}
\label{eq:dispersion_measure_contributions}
    \mathrm{DM}_\mathrm{tot}(\hat{\boldsymbol{x}},z) = \mathrm{DM}_\mathrm{LSS}(\hat{\boldsymbol{x}},z) + \mathrm{DM}_\mathrm{MW}(\hat{\boldsymbol{x}}) + \mathrm{DM}_\mathrm{host}(z)\;,
\end{equation}
where $\mathrm{DM}_\mathrm{LSS}(\hat{\boldsymbol{x}},z)$ is the DM caused by the electron distribution in the large-scale structure, while $\mathrm{DM}_\mathrm{MW}(\hat{\boldsymbol{x}})$ and $\mathrm{DM}_\mathrm{host}(z)$ describe the contributions from the Milky Way and the host galaxy (and its halo),
respectively. Note that \cref{eq:delay_measure} holds in the rest-frame, so the initial host galaxy contribution $\mathrm{DM}_\mathrm{host}$ is observed as $\mathrm{DM}_\mathrm{host}(z) = (1+z)^{-1}\mathrm{DM}_\mathrm{host}$.

Models of the galactic electron distribution predict $\mathrm{DM}_\mathrm{MW} \sim 60 \; \mathrm{pc}\;\mathrm{cm}^{-3}$ \citep{yao_new_2017,platts_data-driven_2020}, and we expect similar values for the host galaxy. For the remainder of this paper, we will assume that the Milky Way contribution can be modelled and subtracted from the measured FRB signal, while $\mathrm{DM}_\mathrm{host}$ is affected by random scatter as discussed in section~\ref{subsec:dispersion_measure_tomography}. The remaining large-scale structure contribution can be written as
\begin{equation}
\label{eq:dispersion_measure_general}
    \mathrm{DM}_\mathrm{LSS}(\hat{\boldsymbol{x}},z) = \int_0^z n_\mathrm{e}({\boldsymbol{x}},z^\prime) \frac{1+z^\prime}{H(z^\prime)}\mathrm{d}z^\prime\;,
\end{equation}
where $H(z)$ is the Hubble expansion rate and the electron density $n_e$ depends on the local matter over-density, $\delta_\mathrm{m}(\boldsymbol{x},z)$:
\begin{equation}
\label{eq:electron_number_density}
    n_\mathrm{e}({\boldsymbol{x}},z) = \frac{\rho_\mathrm{b}({\boldsymbol{x}},z)}{m_\mathrm{p}} = \frac{\bar\rho_\mathrm{b}(z)}{m_\mathrm{p}}\left[1+b_\mathrm{e}({\boldsymbol{x}},z)\delta_\mathrm{m}({\boldsymbol{x}},z)\right] \; ,
\end{equation}
with the mean baryon density $\bar\rho_\mathrm{b}(z)${, the proton mass $m_\mathrm{p}$} and the electron clustering bias $b_\mathrm{e}({\boldsymbol{x}},z)$. Rewriting this expression yields
\begin{equation}
\label{eq:dispersion_measure_specific}
  \!\!  \mathrm{DM}_\mathrm{LSS}(\hat{\boldsymbol{x}},z) = \mathcal{A}\!\int_0^z\!\!\! \mathrm{d}z^\prime \frac{1+z^\prime}{E(z^\prime)}F(z^\prime)[1\!+\!b_\mathrm{e}({\boldsymbol{x}},z^\prime)\delta_\mathrm{m}({\boldsymbol{x}},z^\prime)]\;,
\end{equation}
with
\begin{equation}
    \mathcal{A} \equiv \frac{3H_0^2\Omega_{\mathrm{b}0}\chi_H}{8\uppi G m_\mathrm{p}}\;,
\end{equation}
where
$\chi_H$ {is} the Hubble radius today,
{$E(z) = H(z) / H_0$ is the dimensionless expansion rate},
and $F(z)$ describes the mass fraction of electrons in the IGM. The latter can be written as
\begin{equation}
    \label{eq:ionization_fraction_of_the_igm}
    F(z) = f_\mathrm{IGM}(z)[Y_\mathrm{H}X_{\mathrm{e},\mathrm{H}}(z) + Y_\mathrm{He}X_{\mathrm{e},\mathrm{He}}(z)]\;,
\end{equation}
with $Y_\mathrm{H} = 0.75$ and $Y_\mathrm{He} = 0.25$ being the mass fractions of hydrogen and helium, respectively, and $X_{\mathrm{e},\mathrm{H}}(z)$ and $X_{\mathrm{e},\mathrm{He}}(z)$
{their}
ionization fractions. Hydrogen and helium are {both} fully ionized for $z\lsim 3$ \citep{meiksin_physics_2009,becker_detection_2011} relevant for the surveys considered here and we set $X_{\mathrm{e},\mathrm{H}} =X_{\mathrm{e},\mathrm{He}} = 1$. Finally, $f_\mathrm{IGM}(z)$ is the fraction of electrons in the intergalactic medium{,} which has a slight reshift dependence \citep{shull_baryon_2012}, with $10\%\; (20\%)$ locked up in galaxies at $z \gsim 1.5 \; (\lsim 0.4)$.
Considering the individual contributions to the DM in \cref{eq:dispersion_measure_contributions}, for a source located at redshift unity the large-scale structure $\mathrm{DM}$ is roughly 20 times larger than the galactic component and dominates the signal. {This makes FRBs a unique cosmological probe compared to weak lensing or galaxy clustering, which require a large amount of tracers to overcome shot noise.} The DM measurements on the other hand are cosmic-variance limited on large angular scales even with only a modest number of sources.

\subsection{Primordial non-Gaussianities and Bias}
\label{subsec:PNG}
\label{sec:bias}
The simplest inflationary models predict Gaussian initial conditions, but inflaton self-interactions or couplings to other fields in the early Universe can leave a non-Gaussian imprint on the primordial curvature fluctuations. Here we consider local transformations of the Gaussian field $\phi_\mathrm{G}$, and Taylor-expand the transformation to get
\begin{equation}
    \phi_\mathrm{NG} = \phi_\mathrm{G} + f_\mathrm{NL} \phi_\mathrm{G}^2 + \mathcal{O}(\phi^3_\mathrm{G}) \; ,
\end{equation}
where the non-Gaussianity is characterised by the parameter $f_\mathrm{NL}$ \citep{desjacques_large-scale_2018}. In the peak-background split, the mode coupling caused by $f_\mathrm{NL}$ leads to a scale-dependent modification of the halo collapse dynamics as short (halo-sized) perturbations respond to the coupling to long background modes. The resulting halo bias $b_\mathrm{h}$ is modified at tree-level,
\begin{equation}
    b_\mathrm{h} \rightarrow b_\mathrm{h} + \Delta b_\mathrm{h}^\mathrm{NG}(k, z ) \;,
\end{equation}
by the new non-Gaussian term \citep{slosar_constraints_2008}
\begin{equation}
\label{eq:non_Gaussian_contribution_slosar}
    \Delta b_\mathrm{h}^\mathrm{NG}(k,z) = 3f_\mathrm{NL}\delta_\mathrm{c}(b_\mathrm{h}-1)\frac{\Omega_\mathrm{m0}{H}_0^2}{aT_\phi(k,z)k^2}\;,
\end{equation}
{where $T_\phi(k,z)$ is the potential transfer function.}
The DM fluctuations are caused by diffuse electrons, and since feedback expels gas out of high-density regions, they are anticorrelated with halos on large scales. Therefore they inherit the PNG bias modification, and the electron power spectrum can be written as
\begin{equation}
\label{eq:tree_level_power_spectrum}
P_\mathrm{ee}(k,z) = \Big(b_\mathrm{e}+\Delta b_\mathrm{e}^\mathrm{NG}(k,f_\mathrm{NL})\Big)^2 P^{(1)}_\mathrm{mm}(k,z)
\end{equation}
where $b_\mathrm{h}$ is replaced with the electron bias $b_\mathrm{e}$ in \cref{eq:non_Gaussian_contribution_slosar}, and $P^{(1)}_\mathrm{mm}(k,z)$ is the linear power spectrum of the matter fluctuation obtained using the CLASS code \citep{lesgourgues_cosmic_2011,blas_cosmic_2011}.
Since the transfer function $T_\phi \approx \mathrm{const.}$ at large scales, the PNG modification in \cref{eq:non_Gaussian_contribution_slosar} introduces a characteristic $k^{-2}$ scaling of the bias not easily mimicked by other effects. It also vanishes for every tracer that is on average unbiased.
 
Since feedback pushing gas out of halos is less effective in low density regions, the electron fluctuations are smoothed compared to the underlying dark matter.
This effect can be seen in hydrodynamical simulations, which find $b_\mathrm{e} \approx \mathrm{const.} < 1$ on large scales
\citep{fedeli_clustering_2014,shaw_deconstructing_2012,springel_first_2018,mead_hydrodynamical_2020}. Feedback expels the baryons from halos gradually, so the electron bias has a redshift evolution{, approaching} unity
at high redshifts. As soon as $b_\mathrm{e} \approx 1$, the PNG imprint on the electron distribution vanishes.
We will presents our results in terms of the large-scale electron bias today, $b^0_\mathrm{e}$, and the typical redshift when feedback starts expelling electrons from halos, $z_\mathrm{fb}$, with a linear evolution in between {and}
$b_\mathrm{e}(z>z_\mathrm{fb}) = 1$. Typical values from $N$-body simulations are $b^0_\mathrm{e} = 0.75$ and $z_\mathrm{fb} = 5$ \citep{shaw_deconstructing_2012}, which we will adapt unless mentioned otherwise. Since the strength and modelling of feedback in simulations is still under debate, we present final results for various $b^0_\mathrm{e}$ and $z_\mathrm{fb}$.

\begin{figure}
\begin{center}
\includegraphics[width = 0.45\textwidth]{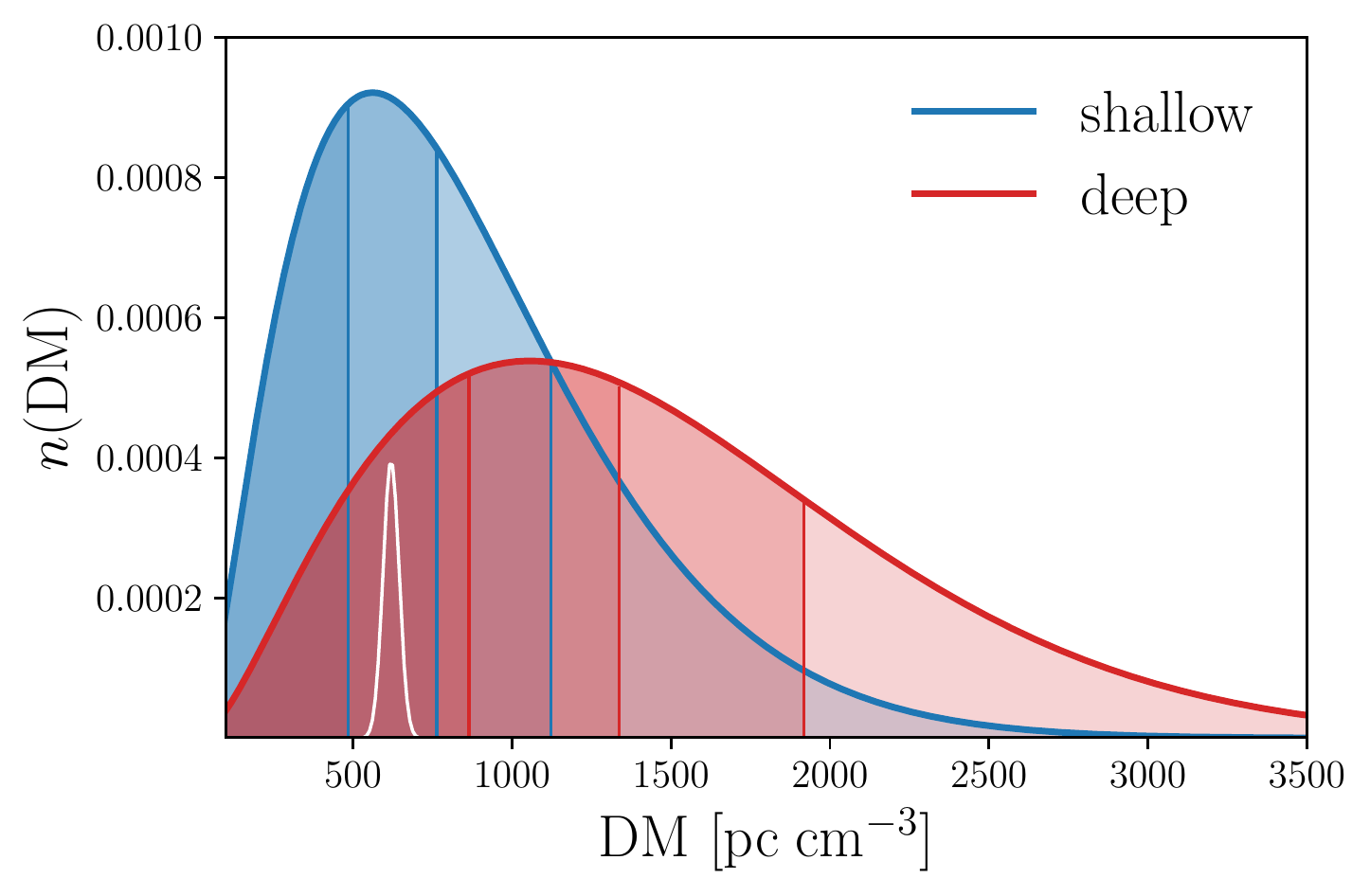}
\caption{Normalized source distribution $n(\mathrm{DM})$ as a function of the DM for two different choices of $\alpha$ \cref{eq:source_redshift_distributiob} with four tomographic bins. The configurations are shown in blue and red for the shallow ($\alpha=3.5$) and the deep ($\alpha=2$) surveys, respectively.
The white curve shows an example for the {DM}
distribution due to line-of-sight fluctuations and the intrinsic host scatter, \cref{eq:photo_z_equi}.}
\label{fig:srd_tomography}
\end{center}
\end{figure}

\subsection{Dispersion measure statistics}
\label{subsec:dispersion_measure_statistics}
In this section we discuss the statistics of fluctuations of the DM. We start by considering deviations from the mean DM contribution at a given redshift:
\begin{equation}
\label{eq:fluctuation_definition}
     \mathrm{DM}_\mathrm{LSS}(\hat{\boldsymbol{x}},z) = \langle  \mathrm{DM}_\mathrm{LSS}\rangle (z) + \mathcal{D}(\hat{\boldsymbol{x}},z)\,.
\end{equation}
$\mathcal{D}(\hat{\boldsymbol{x}},z)$ is the effective DM induced by the fluctuations in the LSS and is given by a weighted line-of-sight integral over the {electron density}.
Given a normalised source redshift distribution $n(z)$, such that $\int\mathrm{d}z\; n(z) = 1$, {and the associated distance distribution $n(\chi) = n(z) \, \mathrm{d}z / \mathrm{d}\chi$,} \cref{eq:dispersion_measure_specific} can be averaged over redshift:
\begin{equation}
    \label{eq:dispersion_measure_averaged}
    \mathcal{D}(\hat{\boldsymbol{x}}) = \int_0^{\chi_H}\!\!\mathrm{d}\chi\; n(\chi) \mathcal{D}\big(\hat{\boldsymbol{x}},z(\chi)\big)\;. 
\end{equation}
Rearranging integration limits yields
\begin{equation}
    \label{eq:averaged_dispersion_measure_fluctuation}
    \mathcal{D}(\hat{\boldsymbol{x}}) = \int_0^{\chi_H}\!\!\mathrm{d}\chi\; W_\mathcal{D}(\chi)\delta_\mathrm{e}
    \big(\hat{\boldsymbol{x}},z(\chi)\big) \;,
\end{equation}
with the averaged weighting function
\begin{equation}
\       \label{eq:averaged_dispersion_measure_fluctuation_weighting_function}
        W_\mathcal{D}(\chi) = W(\chi)\int_\chi^{\chi_H} \mathrm{d}\chi^\prime n(\chi^\prime) \; ,
\end{equation}
and $W(\chi)$ {being}
defined via \cref{eq:dispersion_measure_specific}:
\begin{equation}
    \label{eq:averaged_dispersion_measure_fluctuation_weighting_function_inner}
    W(\chi) = \mathcal{A}\frac{F \big(z(\chi) \big) \big(1+z(\chi) \big)}{E\big(z(\chi)\big)}\left|\frac{\mathrm{d}z}{\mathrm{d}\chi}\right|\;.
\end{equation}
The angular power spectrum of DM correlations for the source distribution is then given by
\begin{equation}
\label{eq:dispersion_measure_angular_power_spectrum}
\begin{split}
    C^{\mathcal{D}\mathcal{D}}&(\ell) =  \;\frac{2}{\uppi}\int\mathrm{d}\chi_1 W_\mathcal{D}(\chi_1)\int\mathrm{d}\chi_2W_\mathcal{D}(\chi_2) \\
    & \times \int k^2\mathrm{d}k\;  \sqrt{P_{\mathrm{ee}}(k,\chi_1)P_{\mathrm{ee}}(k,\chi_2)}j_{\ell}(k\chi_1)j_{\ell}(k\chi_2)\;,
    \end{split}
\end{equation}
where we compute the unequal time-correlator as the geometric mean of the power spectra at the corresponding times.
 While the contribution of the Milky Way can be removed from the total DM signal, the electron distribution of the host galaxy acts as a stochastic source with an intrinsic width of $\sigma^2_\mathrm{host}$. This adds a white noise contribution to the observed spectrum due to the finite amount of sources per solid angle, $\bar{n}$:
\begin{equation}
\label{eq:dispersion_measure_angular_power_spectrum_observed}
    {C^{\mathcal{D}\mathcal{D}}(\ell) \to C^{\mathcal{D}\mathcal{D}}(\ell)}
    + \frac{\sigma^2_\mathrm{host}}{\bar{n}}\;.
\end{equation}
It remains to specify the {observable} source redshift distribution.
We assume that FRBs roughly trace the galaxy distribution and take the following form:
\begin{equation}
\label{eq:source_redshift_distributiob}
    n(z) \propto z^2\exp\left(-\alpha z\right)\;{.}
\end{equation}
To cover a range of possible survey characteristics, in this work we will calculate results for {a shallow survey with $\alpha = 3.5$ and a deeper survey with $\alpha = 2$. Unless stated otherwise we will assume a total number of $5\times10^3$ FRBs for the shallow and $5\times10^4$ FRBs for the deep survey.}

\begin{figure}
\begin{center}
\includegraphics[width = 0.45\textwidth]{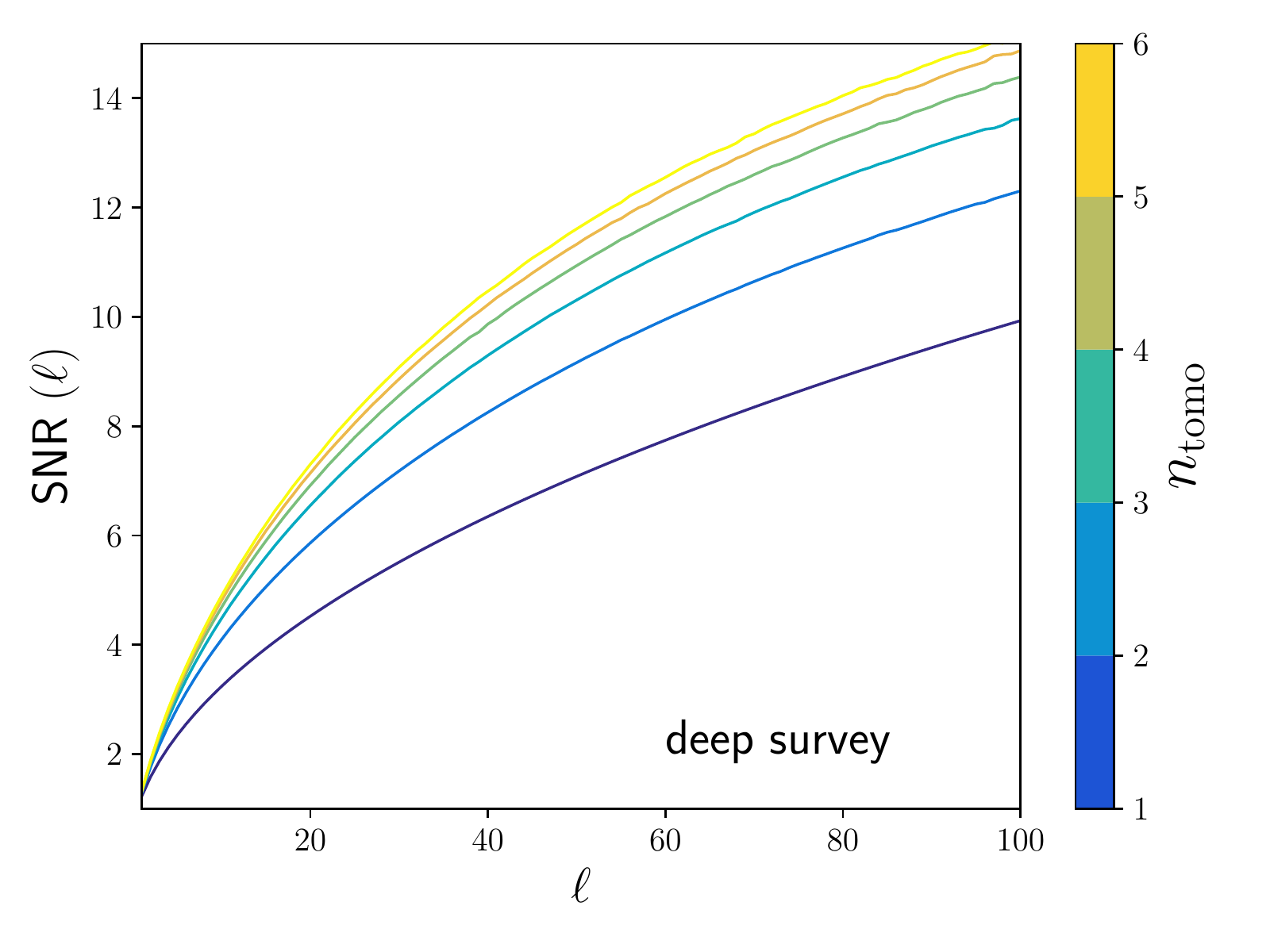}
\caption{Signal-to-noise ratio (SNR) of the deep survey at each multipole for different numbers of tomographic bins. More bins increase the signal significantly, but {also} increase {the} correlation {between bins}, so the gain saturates {around}
$n_\mathrm{tomo} \sim 4$ for the chosen survey characteristics.
}
\label{fig:SNR}
\end{center}
\end{figure}

Redshifts cannot be determined from the FRB observations directly, and need to be inferred by host galaxy identification. The source redshift distribution might therefore not be known precisely. However, we can invert \cref{eq:dispersion_measure_specific} to estimate the underlying FRB redshift distribution directly from the observed DM histogram of all sources. Since the DM for a given redshift fluctuates due to different host galaxy contributions and variations in the electron density along the line of sight, we
marginalise over the associated uncertainties and write the estimated source distribution as
\begin{equation}
\label{eq:observed_srd}
    n(z) = \int \mathrm d \mathrm{DM} \; n(\mathrm{DM}) p(\mathrm{DM} | z) \;.
\end{equation}
Hydrodynamical simulations suggest that the probability distribution $p(\mathrm{DM}|z)$ is close to a Gaussian \citep{jaroszynski_fast_2019}
\begin{equation}
\label{eq:photo_z_equi}
    p(\mathrm{DM}|z) \sim \mathcal{N}(\langle\mathrm{DM}\rangle (z),\sigma^2_{\mathrm{DM}(z)})\;,
\end{equation}
with mean
\begin{equation}
\label{eq:mean_dispersion_measure}
    \langle \mathrm{DM}\rangle(z) = \langle \mathrm{DM}_\mathrm{host} \rangle (z)+ \langle\mathrm{DM}_\mathrm{LSS}\rangle(z) \;,
\end{equation}
and variance
\begin{equation}
\label{eq:sigma_dispersion_measure}
    \sigma^2_{\mathrm{DM}(z)} = \sigma^2_\mathrm{host}(z) + \sigma^2_\mathrm{LSS}(z)\;,
\end{equation}
which consists of the scatter from the host galaxy and a cosmological contribution $\sigma^2_\mathrm{LSS}$. In \citep{mcquinn_locating_2014,jaroszynski_fast_2019}, $\sigma_{\mathrm{LSS}(z)}/\mathrm{DM}_\mathrm{LSS}(z)$ was found to be $13\%$ at $z=1$ and $7\%$ at $z=3$. We adopt these values with a linear dependence on redshift for our model. For the noise introduced by the host galaxy, we assume  $\langle\mathrm{DM}_\mathrm{host} \rangle (z) = 50 \mathrm{pc}\;\mathrm{cm}^{-3} (1+z)^{-1}$ and $\sigma_{\mathrm{host}}(z) = 50 \mathrm{pc}\;\mathrm{cm}^{-3}(1+z)^{-1}$.

Note that the conversion from DM into redshift or distance is only possible due to the relatively small relative fluctuations in the total DM for each FRB. For our forecast it also requires a fixed cosmology, but when analysing real data the conversion can be performed self-consistently together with the cosmological fit. The shape of $p(\mathrm{DM}|z)$ might also turn out to be more complicated as more FRBs are discovered, but a more general distribution can easily be accommodated within the same framework.

\begin{figure*}
\begin{center}
\includegraphics[width = 0.95\textwidth]{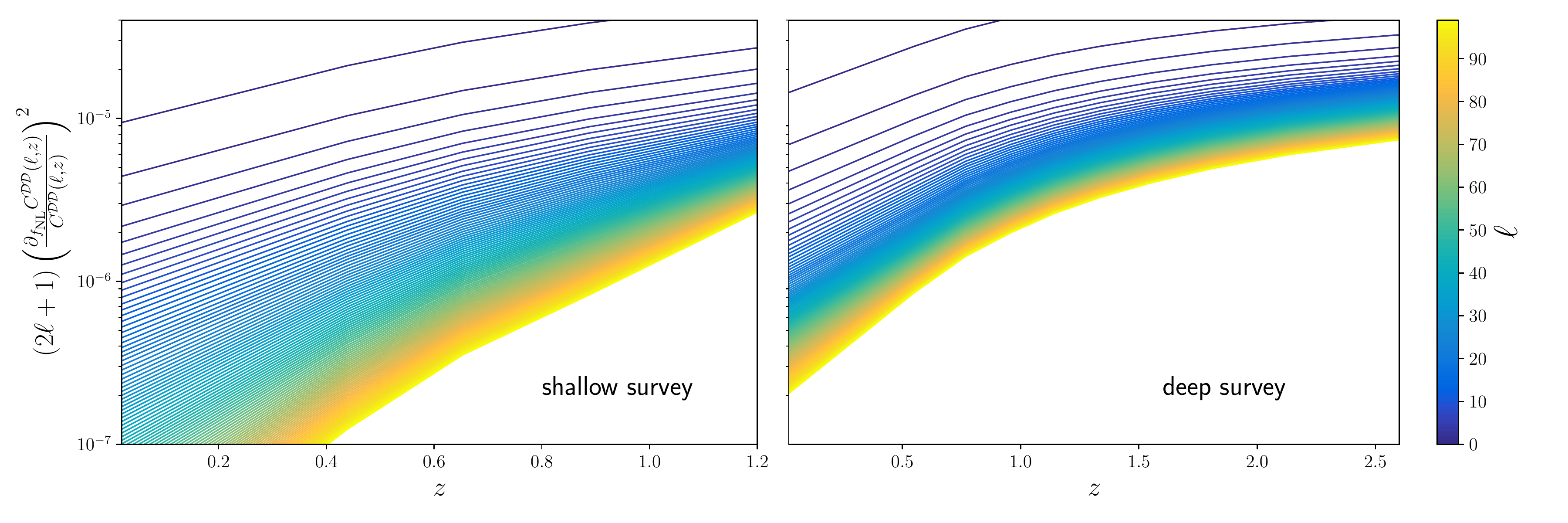}
\caption{Differential sensitivity with respect to $f_\mathrm{NL}$ as a function of redshift and multipole (colour bar). On the left (right) the results for the shallow (deep) survey are shown.
Note that different redshifts are correlated and are just shown to illustrate from which scales and redshifts the signal originates.
}
\label{fig:sensitivity}
\end{center}
\end{figure*}

\subsection{Tomographic dispersion measure statistic}
\label{subsec:dispersion_measure_tomography}
Since the underlying redshift distribution of FRBs is assumed to be broad, the angular DM correlations in \cref{eq:dispersion_measure_angular_power_spectrum} are sourced by a wide range of scales. This makes it challenging to isolate the scale-dependent bias modification caused by primordial non-Gaussianity.

To improve the sensitivity, we divide the source redshift distribution into $n_\mathrm{tomo}$ tomographic redhsift bins $i$.
In order to do this, each source is assigned to a redshift bin by inverting \cref{eq:mean_dispersion_measure} for the mean DM.
Introducing redshift bins $z_i$, we modify the weight function, \cref{eq:averaged_dispersion_measure_fluctuation_weighting_function}, so that the weight function in the $i$-th bin is given by
\begin{equation}
\label{eq:tomographic_weight}
    G^i_\mathcal{D}(\chi) = W(\chi)\int_{\mathrm{max}(\chi,\chi_i)}^{\chi_{i+1}}\mathrm d\chi^\prime\;n(\chi^\prime)\;,
\end{equation}
where $\chi_i$ is the comoving distance boundary of the $i$-th bin, $\chi_i=\chi(z_i)$. The angular power spectra consist of $n_\mathrm{tomo}(n_\mathrm{tomo}+1)/2$ independent numbers $C^{\mathcal{D}\mathcal{D}}_{ij}(\ell)$ for each multipole $\ell$. We split the sample into different DM bins such that each contains the same amount of sources. This increases the shot noise contribution, \cref{eq:dispersion_measure_angular_power_spectrum_observed}, by a factor of $n_\mathrm{tomo}$
\begin{equation}
    \label{eq:tomographic_shot_noise}
    \frac{\sigma^2_\mathrm{host}}{\bar{n}} \to \frac{n_\mathrm{tomo}\sigma^2_\mathrm{host}}{\bar{n}}\delta_{ij}\;,
\end{equation}
since different tomographic bins are uncorrelated in their noise properties.

\Cref{fig:srd_tomography} shows the source distribution as a function of DM for both surveys. In red (blue) we depict the corresponding bins in DM for the shallow (deep) survey such that on average each bin contains an equal number of FRBs{, assuming four bins in total}. The white curve shows the typical scatter around the average DM due to the host galaxy contribution and the LSS, which is small compared to the chosen size of the tomographic bins and has a negligible effect on the spectra. This changes as more and more bins are introduced. While one can in principle propagate the DM uncertainty rigorously by working with the full 3-dimensional rather than the (binned) projected DM field \citep{masui_dispersion_2015},
for this forecast we adopt a simple binned approach.

To demonstrate the advantages of the tomographic approach, \cref{fig:SNR} shows the signal-to-noise ratio (SNR) of the angular DM power spectrum measurement at each multipole for the deep survey for a varying number of bins. We calculate the SNR as
\begin{equation}
    \mathrm{SNR}^2(\ell) = \frac{2\ell +1}{2}\mathrm{tr}\left(\boldsymbol{S}\boldsymbol{C}^{-1}\boldsymbol{S}\boldsymbol{C}^{-1}\right)\;,
\end{equation}
where $\boldsymbol{S}$ is the matrix containing the tomographic spectra without the shot noise, and $\boldsymbol{C} = \boldsymbol{S} + \boldsymbol{N}$ is the total noise contribution including cosmic variance and shot noise. Adding tomographic bins increases the SNR of the measurement, but also increasingly correlates the individual bins. Therefore adding more and more bins has diminishing returns on the SNR. As can be seen, the SNR already starts to saturate at $n_\mathrm{tomo}\approx 4$, which we adopt for all further results. Also note that the SNR is still well above the shot noise at $\ell =100$.

One can also study the clustering of FRBs directly. Since they reside in galaxies, this is analogous to galaxy clustering survey. While cross-correlations between the FRBs and galaxy positions can be used to infer the source-redshift distribution statistically \citep{rafiei-ravandi_characterizing_2019}, the FRB auto-correlation is dominated by shot noise. FRB clustering would require at least two orders of magnitude more sources to be competitive with the constraints from the DM correlations alone.

\section{Results}
\label{sec:results}
\subsection{Fisher matrix}
For our forecasts we assume Gaussian likelihoods and can thus express the probability of finding a set of modes $\{\mathcal{D}_{\ell m,i}\}$ given model parameters $\bm{\theta}$ (which in our case is just $f_\mathrm{NL}$) {as}:
\begin{equation}
p\left(\left\{\mathcal{D}_{\ell m,i}\right\}\big|\bm{\theta}\right) \propto \prod_\ell \left(\mathrm{det}\left(\bm{C}_\ell^{-1}\right) \exp\left[\bm{\mathcal{D}}_{\ell m}^\dagger \boldsymbol{C}_\ell^{\mathstrut -1}\bm{\mathcal{D}}^{\mathstrut}_{\ell m}\right]\right)^{2\ell+1}\!\!\!,
\end{equation}
where we combined all redshift bins into a vector $\bm{\mathcal{D}}_{\ell m}$. Their covariance is given by $\boldsymbol{C}_\ell = \left\langle \bm{\mathcal{D}}^{\mathstrut}_{\ell m}\bm{\mathcal{D}}_{\ell m}^\dagger \right\rangle$, {averaging}
over all possible realizations of the data. The components of the covariance are thus given by \cref{eq:dispersion_measure_angular_power_spectrum_observed} in all tomographic bins. The Fisher matrix can be written as \citep{tegmark_karhunen-loeve_1997}:
\begin{equation}\label{eq:fisher_matrix_gaussian_power}
F_{\alpha\beta} = f_\mathrm{sky}\!\!\!\sum_{\ell = \ell_\mathrm{min}}^{\ell_\mathrm{max}} \frac{2\ell +1}{2} \mathrm{tr}\left(\bm{\hat{C}}_\ell^{{\mathstrut}-1}\partial^{\mathstrut}_\alpha\bm{C}^{\mathstrut}_\ell \bm{\hat{C}}_\ell^{{\mathstrut}-1}\partial^{\mathstrut}_\beta\bm{C}^{\mathstrut}_\ell\right)\bigg|_{\bm{\theta} = \bm{\theta}_0} \;,
\end{equation}
where $\partial_i$ is the derivative with respect to the $i$-th model parameter. The fiducial cosmology $\bm{\theta}_0$ is fixed to \citet{planck_collaboration_planck_2018} and thus has no PNGs. For the sky fraction we choose $f_\mathrm{sky} = 0.9$, accounting for a possible obstruction by the galactic disk. Multipoles are collected up to $\ell_\mathrm{max} = 100$ since there will be no PNG imprint on higher multipoles. We adopt a flat prior on $f_\mathrm{NL}$ for all results.

\begin{figure*}
\begin{center}
\includegraphics[width = 0.95\textwidth]{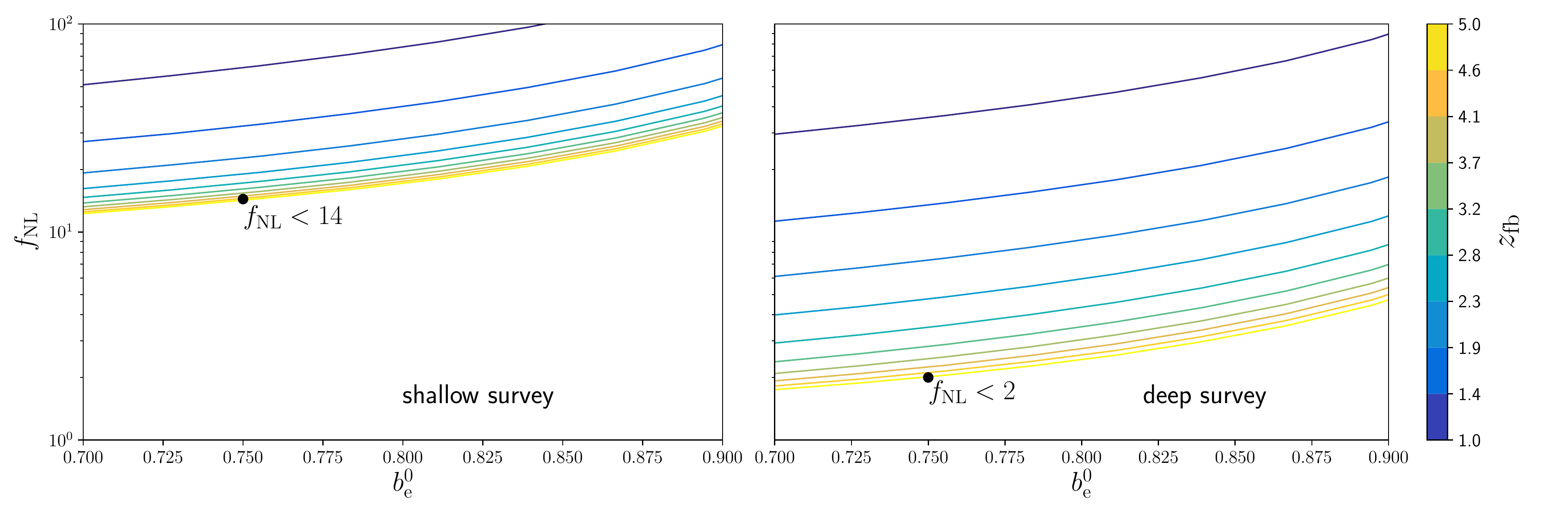}
\caption{$1 \sigma$ limits on $f_\mathrm{NL}$ for the shallow (left) and the deep (right) survey as a function of the bias model. The horizontal axis shows the electron bias at redshift zero. {The colour scale denotes the redshift $z_\mathrm{fb}$}
below which feedback and gas dynamics influence the clustering properties of electrons on large scales. The black dot corresponds to the values found in \citep{shaw_deconstructing_2012}.}
\label{fig:fnl}
\end{center}
\end{figure*}

\subsection{Forecast constraints on PNGs}
Here we present our expected constraints from the DM angular power spectrum on local PNGs parameterised by $f_\mathrm{NL}$. Results are shown in terms of one sigma upper bounds on $f_\mathrm{NL}$ by virtue of the Cram\'{e}r-Rao bound, $\sigma_{f_\mathrm{NL}} = 1/\sqrt{F_{f_\mathrm{NL}f_\mathrm{NL}}}$, with ${F_{f_\mathrm{NL}f_\mathrm{NL}}}$ given by \cref{eq:fisher_matrix_gaussian_power}.

\Cref{fig:sensitivity} presents the sensitivity of $C^{\mathcal{DD}(\ell, z)}$ to the PNG signal in units of the noise as a function of multipole $\ell$ (colour bar) and redshift $z$ for the shallow (left) and the deep survey (right). Since, following \cref{eq:non_Gaussian_contribution_slosar}, the largest effect occurs on large physical scales, the angular spectrum shows the largest response at small multipoles and high redshifts which have more contributions from small $k$. Note that we only show the auto-correlation at a given $z$, but different redshifts are correlated due to the line-of-sight integration.

In \cref{fig:fnl} we show the resulting constraints on $f_\mathrm{NL}$ for various bias assumptions
for both surveys. We vary the bias at redshift zero, $b^0_\mathrm{e}$, and the redshift where the bias becomes unity, $z_\mathrm{fb}$. Typical values for the bias parameters from simulations are marked with a black dot. The deep survey increases the constraints by roughly one order of magnitude due to its lower shot noise and larger signal strength. Furthermore the longer integration path in the deep survey picks up larger contributions from long wavelengths and
increases the ability to observe modifications which occur at the scale of the comoving Hubble radius (compare \cref{eq:non_Gaussian_contribution_slosar}. {For comparison, the effective FRB survey volume is between 250 and 370$\;\mathrm{Gpc}^3$ and thus more than one order of magnitude larger than that of BOSS \citep{giannantonio_improved_2014}}. 
Increasing $b^0_\mathrm{e}$ or decreasing $z_\mathrm{fb}$ decreases the constraints on $f_\mathrm{NL}$ since the electrons will become less biased. On the other hand a decreasing $b^0_\mathrm{e}$ boosts the signal of PNGs. However, it also decreases the overall signal, leading to a noisy measurement. For $b^0_\mathrm{e}\to 1$ the constraints diverge since there is no PNG signal in the electron distribution.
In our forecast, the bias is derived from numerical simulations. It is, however, also possible to determine the bias directly from observations. This can for example be done by cross-correlating the DM with probes of the total matter density such as weak lensing.

\Cref{fig:fnl_nfrb} presents the constraints on $f_\mathrm{NL}$ as a function of the total number of FRBs available in both surveys. For ${\sim 10^3}$ FRBs, limits from both surveys are comparable to current constraints from the LSS \citep{giannantonio_constraining_2012}. The shallow survey becomes competitive with CMB observations when ${\sim 4 \times 10^4}$ FRBs are available, while the deep survey requires {less than $10^4$}
objects.

The key result is that we expect tight limits on primordial non-Gaussianity from FRBs, and a sensitivity of ${f_\mathrm{NL}}\sim \mathcal{O}(1)$ can be achieved. This confirms that DM clustering is in principle competitive with other probes such as galaxy clustering, cosmic shear or the cosmic infrared background \citep{lsst_science_collaboration_lsst_2009,laureijs_euclid_2011,giannantonio_constraining_2012,tucci_cosmic_2016}. One of the largest advantages of FRBs is the low shot noise component since the cosmological signal dominates the intrinsic host contribution for distant sources by an order of magnitude. {For both}
galaxy clustering and cosmic shear, millions of sources are needed to detect the signal, while a few thousand FRBs are already sufficient. This also suggests {that} the study of higher-order correlators such as the bispectrum is promising.

Furthermore, FRBs only {suffer from}
small foreground {contamination} if compared to the cosmic infrared background, where the signal on large angular scales crucial for the detection of PNGs is dominated by galactic dust. The major challenge is that the electron distribution, i.e.~the clustering bias $b_e$, needs to be known quite well. In our analysis we did not marginalize over {the} uncertainty in the model of the electron bias, but presented results for various different values.
{Lastly,}
we would like to stress that our analysis does not rely on the availability of redshifts of the FRBs since the uncertainties on the redshift estimate on the basis of the DM measurements are taken into account.

\section{Conclusion}
\label{sec:conclusion}
In this paper we studied the potential of FRB observations to detect signals from PNGs by exploiting the scale-dependent bias on large scales induced by the non-linear parameter $f_\mathrm{NL}$. We modelled the DM angular correlation function by assuming that the electrons are slightly less clustered compared to the total matter due to {astrophysical} feedback expelling them from overdense regions. We assumed that the electron bias approaches unity with redshift and they stay unbiased for higher redshifts. This sets a redshift $z_\mathrm{fb}$ where feedback in galaxies starts to drive electrons out of dark matter halos. {Consequently, at times $z < z_\mathrm{fb}$, electrons show}
the well known scale-dependent bias on very large scales as induced by local PNGs. Although the effect is in principle weaker than for other probes, such as galaxy clustering, since the electrons are {distributed} much more diffuse{ly}, the constraints can still be competitive due to the small shot noise and the large survey volume and low foreground levels on large scales. 

We summarize our findings as follows:

\begin{figure}
\begin{center}
\includegraphics[width = 0.45\textwidth]{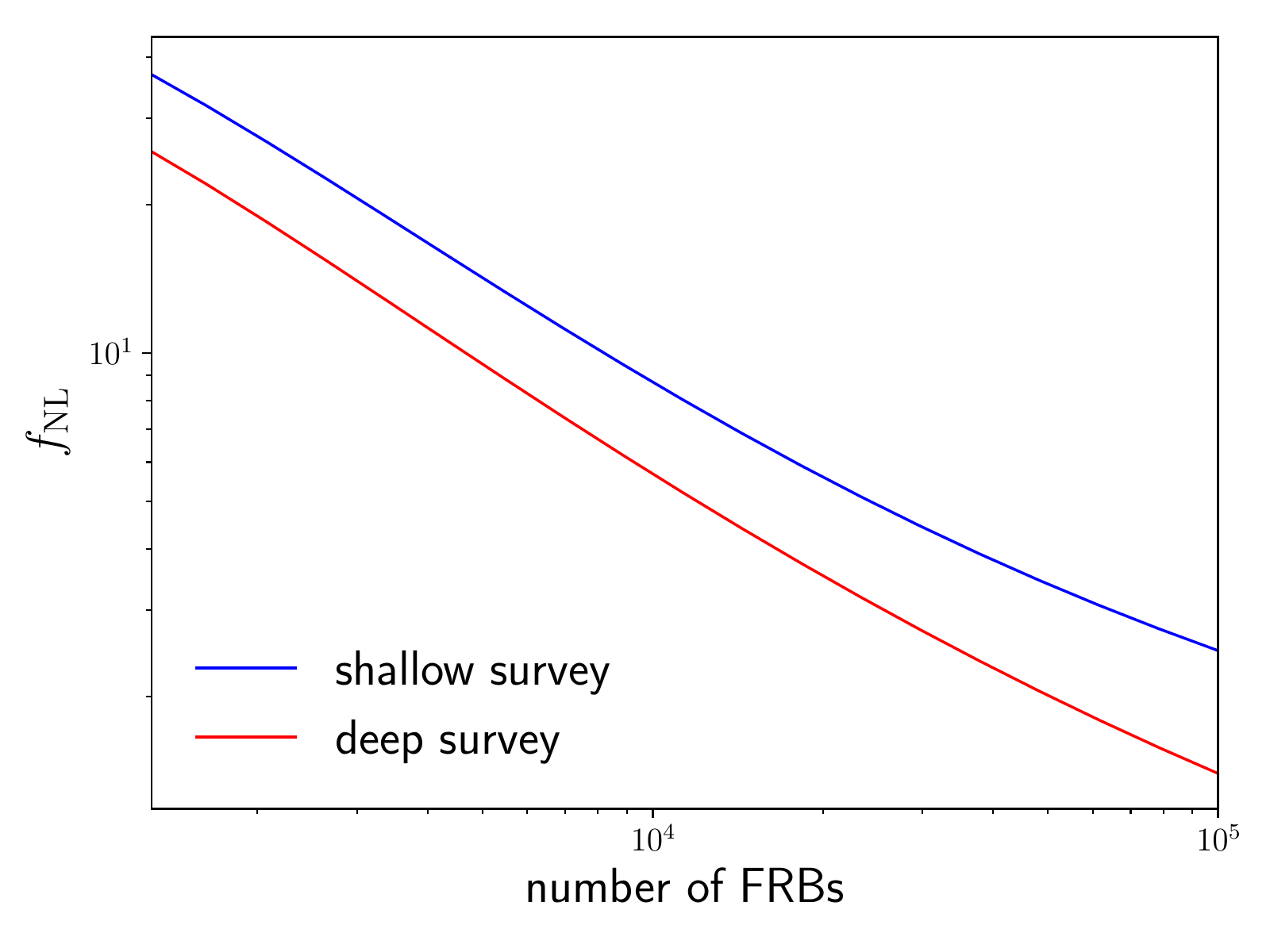}
\caption{$1 \sigma$ constraints on $f_\mathrm{NL}$ for the shallow (blue) and the deep (red) survey as a function of the total number of FRBs observed to sample the DM.
}
\label{fig:fnl_nfrb}
\end{center}
\end{figure}

$(i)$ The DM angular power spectrum is easily measurable in the near future with only a few $10^3$ FRBs due to the strength of the signal compared to its intrinsic fluctuation as given by the host galaxy contribution. In particular, the angular power spectrum dominates the noise contribution up to $\ell \sim 100$. We note that DM correlations are far more promising than the clustering of FRBs themselves, which requires several orders of magnitude more sources to overcome the larger shot-noise.
    
$(ii)$ Since the relative fluctuations are small compared to unity one can reasonably map between the DM and redshift, allowing for a tomographic analysis of the DM even when redshift information of the individual FRBs are not directly available due to lacking host galaxy association.

$(iii)$ PNGs of the local type, as parameterised through $f_\mathrm{NL}$, increase the power of the DM angular power spectrum by a few percent for $f_\mathrm{NL} \sim \mathcal{O}(1)$.

$(iv)$ Constraints on $f_\mathrm{NL}$ vary depending on the survey setting {and the precise influence of astrophysical feedback mechanisms on the large-scale electron distribution}. However, it is possible to measure the amplitude down to $f_\mathrm{NL} \sim \mathcal{O}(1)$, putting tight constraints on more complex models of inflation in the not so distant future. We also find that {using} a tomographic analysis {is}
key to unlocking the sensitivity on $f_\mathrm{NL}$. 

In conclusion, the analysis of future FRB observations is a promising tool to test cosmology and in particular inflation. While more direct probes of the galaxy distribution, such as SPHEREX \citep{dore_cosmology_2014}, can lead to even stronger constraints on $f_\mathrm{NL}$, FRBs can provide excellent validation of the results with different systematics.

\begin{acknowledgments}
The authors thank Amanda Weltman for interesting discussions about FRBs and helpful comments on an early version of this paper. We would like to thank the organisers of the CosmoCon$\beta$ -- Cosmology from Home conference for making these interactions possible in difficult times. RR and RL thank Vincent Desjacques and Adi Nusser for insightful discussions. 
RR acknowledges support by the Israel Science Foundation (grant no. 1395/16 and 255/18). SH acknowledges support from the Vetenskapsr\r{a}det (Swedish Research Council) through contract No.\ 638-2013-8993 and the Oskar Klein Centre for Cosmoparticle Physics. RL acknowledges support by a Technion fellowship.
\end{acknowledgments}

\bibliographystyle{apsrev4-2}
\bibliography{MyLibrary}

\end{document}